# A systematic review of research on large language models for computer programming education

**Meina Zhu**

Learning Design and Technology, Wayne State University, Detroit, MI, the US;

meinazhuiu@gmail.com

**Lanyu Xu**

Computer Science and Engineering Department, Oakland University, Rochester, MI, the US;

lxu@oakland.edu

**Barbara Ericson**

Information, School of Information and Electrical Engineering and Computer Science, College of Engineering, University of Michigan, Ann Arbor, MI, the US; barbarer@umich.edu

**Corresponding Author:**

Meina Zhu

Learning Design and Technology

Wayne State University

365 Education Building

5425 Gullen Mall

Detroit, MI, 48202, USA

Email: meinazhuiu@gmail.com

**A systematic review of research on large language models for computer programming education**

**Abstract**

**Background and Context:** Given the increasing demands in computer programming education and the rapid advancement of large language models (LLMs), LLMs play a critical role in programming education.

**Objective:** This study provides a systematic review of selected empirical studies on LLMs in computer programming education, published from 2023 to March 2024.

**Method:** The data for this review were collected from Web of Science (SCI/SSCI), SCOPUS, and EBSCOhost databases, as well as three conference proceedings specialized in computer programming education. In total, 42 studies met the selection criteria and were reviewed using methods, including bibliometric analysis, thematic analysis, and structural topic modeling.

**Findings:** This study offers an overview of the current state of LLMs in computer programming education research. It outlines LLMs' applications, benefits, limitations, concerns, and implications for future research and practices, establishing connections between LLMs and their practical use in computer programming education. This review also provides examples and valuable insights for instructional designers, instructors, and learners. Additionally, a conceptual framework is proposed to guide education practitioners in integrating LLMs into computer programming education.

**Implications**: This study suggests future research directions from various perspectives, emphasizing the need to expand research methods and topics in computer programming education as LLMs evolve. Additionally, future research in the field should incorporate collaborative, interdisciplinary, and transdisciplinary efforts on a large scale, focusing on longitudinal research and development initiatives.

**Keywords:** Large language models; artificial intelligence; ChatGPT; computer programming education; systematic review

## Introduction

Computer programming education is important in higher education due to its role in fostering students' computational thinking skills (Stehle & Peters-Burton, 2019; Sun et al., 2024), problem-solving and critical thinking skills (Mathew et al., 2019; Wang et al., 2017), which are essential in many professional contexts. However, research indicates that students may face various obstacles in programming learning (Looi et al., 2018), including struggles to create and implement effective plans in their computer programming learning (Kazerouni et al., 2019), a lack of relevant technical skills and barriers to accessing essential resources (Bau et al., 2017; Tom, 2015). This can be a frustrating experience (Becker et al., 2018) that hinders student learning and may lead to plagiarism (Hellas et al., 2017). Therefore, providing support to the students' programming learning is crucial (Lu et al., 2017).

A variety of technologies have played an important role in supporting computer programming learning. Intelligent tutoring systems (ITS) offer adaptive feedback based on students' learning performance; however, ITS typically has pre-determined tasks (Crow et al., 2018). Recently, automated hint-generation systems have been used to provide hints for various programming exercises (McBroom et al., 2021). These systems typically make decisions based on students' historical learning performance data to identify appropriate approaches for each exercise solution (Keuning et al., 2018; Mahdaoui et al., 2022).

Since the release of ChatGPT 3.5, a natural language processing model developed by OpenAI in November 2022, large language models (LLMs) have attracted the interest of researchers and practitioners in computer programming education. LLM-powered tools such as ChatGPT, GitHub Copilot, and Bard can generate code automatically, potentially influencing students' learning in computer programming (Prather et al., 2023). The easy accessibility of LLM tools indicates the need for updated research that examines the capabilities and implications of the advanced models in programming education. Although some empirical studies have investigated how LLMs can be leveraged in computer programming education, no systematic review examines their usage and impact on students' computer programming education. Therefore, this systematic review aims to address this gap by exploring the influence of LLMs in computer programming education, covering their use, benefits, limitations, concerns, and implications for future research and practices.

## Background on LLMs

### *Large language models*

LLMs are often based on deep learning transformer architecture. These models frequently outperform previous models in most generation tasks. LLMs are typically pre-trained by the developers and made accessible by other users. Users interact with an LLM through prompts, which are natural language inputs that direct the model to produce the expected output. However, users typically do not always understand the internal mechanics of LLMs, leading to the development of methods for constructing effective prompts, which is prompt engineering (Denny et al., 2023; Liu et al., 2023).

Besides natural language text generation, LLMs can also generate code, such as ChatGPT (Sun et al., 2024), GitHub Copilot powered by Open AI Codex (Prather et al., 2023; Wermelinger, 2024), Deepmind AlphaCode (Li et al., 2022), CodeBERT (Feng et al., 2020), etc. Although most of those LLMs are targeted at professionals, ChatGPT and GitHub Copilot are free for students to access, with minimal barriers to use for novice programming learners.

These models have demonstrated a surprising ability to generate functional code. For example, Codex was found to solve around 80% of past introductory programming exam problems, a level of performance that placed Codex in the top quartile of students in the course (Finnie-Ansley et al., 2022). A subsequent study found that Codex performed well in Data Structures and Algorithms exams (Finnie-Ansley et al., 2023).

### *Large language models for computer programming education*

Although there is research on LLMs' educational applications, studies focusing specifically on computer programming are limited, particularly high-quality peer-reviewed articles. The few existing research on using LLMs in computer programming education primarily focus on the following aspects. First, research may focus on LLMs' performance in computer programming and their effectiveness as a programming assistant. For example, Tian et al. (2023) explored ChatGPT's potential in this role, finding it effective in addressing programming challenges as well as highlighting its limitations in comprehending broader problems. Similarly, Surameery and Shakor (2023) analyzed ChatGPT's approaches and constraints in debugging programming. Kashefi and Mukerji (2023) examined ChatGPT's capabilities in programming numerical algorithms. Moreover, Wermelinger (2023) found that Copilot may have difficulty following instructions when solving programming exercises. Furthermore, LLMs may struggle with computational thinking tasks (Bellettini et al., 2023).

Second, the research explores the ways learners can benefit from LLMs in learning programming as learning assistance to enhance their programming skills and academic performance. For instance, studies examine opportunities for scientists and engineers to leverage AI tools like ChatGPT to enhance programming skills (Guo, 2023) and evaluate the methods and assess ChatGPT's potential and limitations in fostering programming skill development (Rahman & Watanobe, 2023). Additionally, studies explore the effects of programming education with ChatGPT on students' computational thinking, programming self-efficacy, and learning motivation (Yilmaz & Yilmaz, 2023a).

Third, research has investigated how LLMs can be used to develop and improve educational resources such as programming exercises (Sarsa et al., 2022) and programming error messages (Leinonen et al., 2023). For example, LLMs can generate a variety of programming exercises with specific themes and concepts that personalize novice learners' learning (Sarsa et al., 2022). Additionally, LLMs can enhance programming error messages by providing explanations and guidance on correcting errors, which potentially benefits instructors and learners (Leinonen et al., 2023).

Although recent research has explored the use of LLMs in computer programming education, there remains a gap in systematically reviewing the current state of this field. Conducting a systematic review is needed to establish a baseline by understanding the current landscape, informing future research regarding research gaps and opportunities, and providing evidence-based insights on research and practices. To address this gap, the study conducted a systematic review of recent empirical research on LLMs in computer programming education, focusing on their usage, benefits, limitations, concerns, and implications for future research and practice. This review study addresses the following questions:

1. What is the current state of research publications on the use of LLMs in computer programming education?
2. How are LLMs being used in computer programming education?
3. What are the benefits of using LLMs in computer programming education?

4. What limitations and concerns are associated with the use of LLMs in computer programming education?
5. What are the implications of existing research for future studies and practice related to LLMs in computer programming education?

## Methods

This systematic review adopted the principles outlined in the Preferred Reporting Items for Systematic Reviews and Meta-Analyses (PRISMA) guidelines (Page et al., 2021). In the following sections, researchers will detail the systematic review procedures used in this study.

### ***Data collection***

#### *Databases*

The databases searched in this systematic review include widely recognized databases in education and LLMs: Web of Science (SCI/SSCI), SCOPUS, and EBSCOhost. Web of Science (SCI/SSCI) was selected as it has a high reputation and includes journals in both the Science Citation Index (SCI) and the Social Science Citation Index (SSCI). SCOPUS and EbscoHost were selected as they are well-recognized databases for journal articles related to social science and education. To ensure the quality of the articles reviewed in this systematic review, only empirical studies from peer-reviewed journals in the three databases were included in this study.

Moreover, this systematic review included three high-quality conference proceedings from the Technical Symposium on Computer Science Education (SIGCSE), the Conference on Innovation and Technology in Computer Science Education (ITiCSE), and the International Computing Education Research (ICER) conference. These conference proceedings were chosen for their focus on computer programming education, affiliation with the Association of Computing Machinery (ACM), and high impact factors.

Following the initial article search and screening, a snowballing approach based on predefined selection criteria was implemented. This method allows researchers to explore additional articles from references that may not have been captured by the original search strings used in databases (Wholin, 2014). These strategies were adopted to ensure comprehensive coverage of relevant literature for the systematic review.

*Searches and selections*

A structured literature search strategy was used in the diverse databases mentioned above, using keywords customized to the specific requirements of each database. In the searches conducted in the three databases and the manual three conference proceeding searches, specific keywords related to LLMs and computer programming education were used to search for article abstracts. These search keywords were "chatgpt," "bard," "claude," "github copilot," "large language model," "programming learning," "programming teaching," "programming education," "computer science education," and "software engineering education."

To ensure a systematic and consistent approach, inclusion and exclusion criteria were established and applied to the literature selection process (see Table 1). The reason to specifically focus on LLMs rather than diverse AI tools is because LLMs have a significant influence on programming with their function of code generation, etc. Second, this study specifically excluded K-12 education as these computer programmings were primarily taught in higher education settings. Third, only empirical studies involving data were included, as this study focuses on LLMs using practices in programming education. Fourth, only peer-reviewed journals through three databases and three specific conference proceedings (i.e., SIGCSE, ITiCSE, and ICER) were included to ensure the quality of the reviewed studies from the initial stage. Fifth, the articles reviewed were primarily published between 2023 and March 2024 (the time that the search was conducted), as the attention and popularity of using LLMs for programming were primarily since November 2022, when ChatGPT 3.5 was released by OpenAI. Last, due to researchers' language limitations, the reviewed articles primarily focus on English publications.

**Table 1** A summary of the inclusion and exclusion criteria.

| **Inclusion criteria** | **Exclusion criteria** |
|---|---|
| Research must involve using LLMs in computer programming education. | Research not related to LLMs or not in computer programming education was excluded. |
| Research must focus on higher education settings. | Research on LLMs use in K-12 or other settings was excluded. |
| Research must be empirical studies. | Review articles and theoretical or conceptual papers were excluded. |
| Research must be published in peer-reviewed journals from three databases or conference proceedings from SIGCSE, ITiCSE, and ICER. | Articles published as book chapters, magazines, reports, dissertations, preprints, white papers, cases, self-publication, etc., were excluded. |
| The publication should be from 2023- March 2024. | The publication before 2022 were excluded |
| Articles must be published in English. | Non-English publications were excluded. |

The screening process included the following steps: (1) removing the duplicated articles based on the titles, (2) deleting articles that did not meet the inclusion criteria based on the title and abstracts, (3) deleting articles that did not meet the inclusion criteria based on the full-text reading, (4) using the snowballing approach to go back and forward reference search to further locate the articles, and (5) extracting data from the final articles (see Figure 1). Zotero was used for article management and screening.

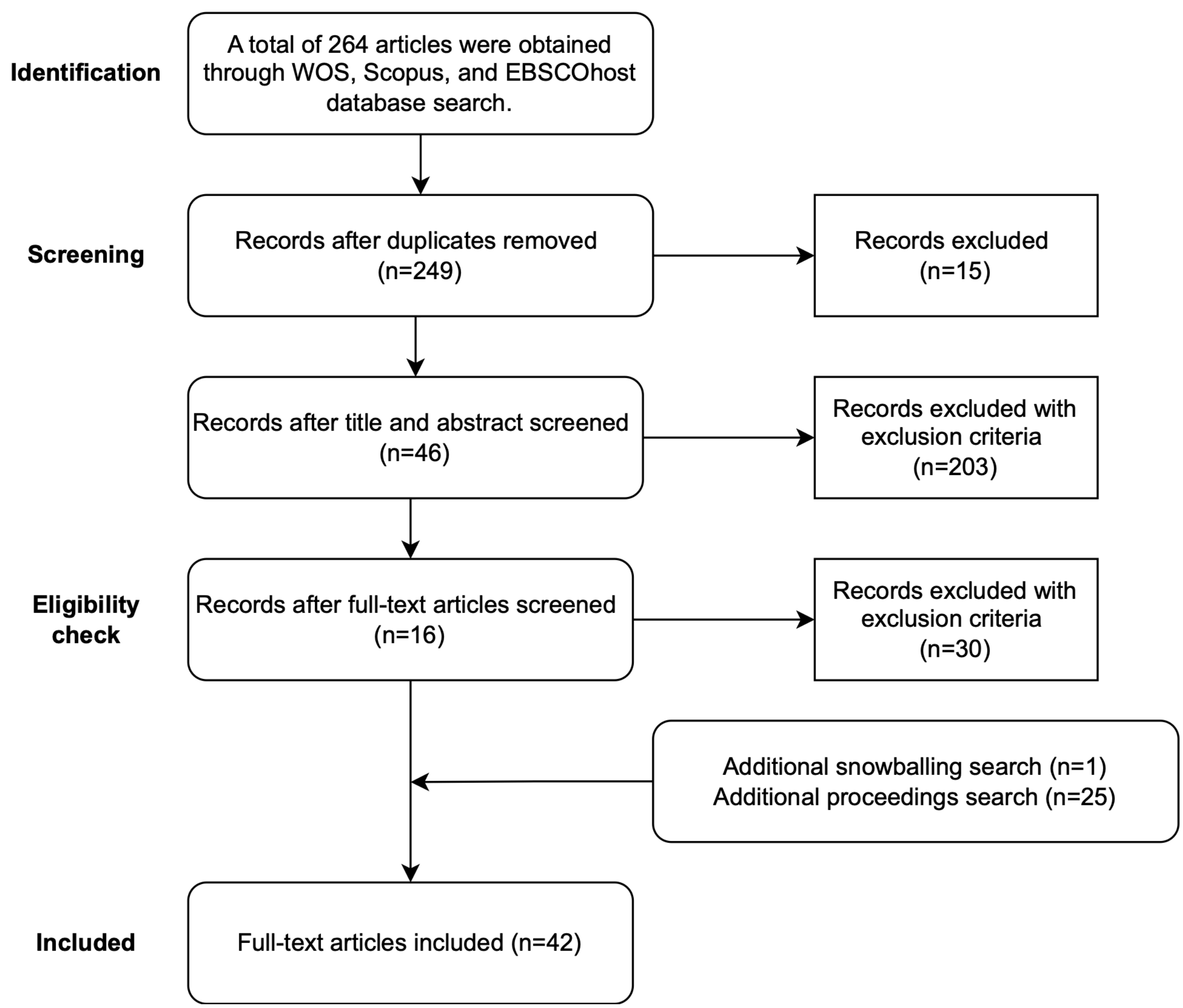

**Fig. 1.** Research and selection procedure based on the PRISMA flowchart

The initial searches of the three databases yielded 264 articles, out of which 15 duplications were excluded. The screening of the title and abstract yielded 46 potential papers for review. Then, the researchers screened the full text, excluded articles that had low research quality and did not meet inclusion criteria, and identified 16 eligible articles. For instance, if the studies did not have empirical data, we did not include them in the review. Through the snowballing journal searches, one additional article that met the inclusion criteria was located. Moreover, through further search from the conference proceedings of ITiCSE, SIGCSE, ICER, 25 proceeding articles were located. Thus, a total of 42 eligible articles were selected for the systematic review.

***Data analysis***

The data analysis was conducted through multiple phases. Bibliometric analyses (e.g., Thelwall, 2008), thematic analysis (Braun & Clarke, 2019), and topic modeling (Roberts et al., 2014) were conducted on all eligible research articles. Major codes included bibliometrics of the publication (e.g., title, year of publication, name of the journal or conference proceedings), country of author affiliation, number of participants, the learning environment, learning activities, learning outcomes, educational benefits, challenges, and concerns, implications for future research and practices, diversity, equity, and inclusiveness, etc.

The first author analyzed the bibliometrics of the articles accordingly. Based on Campbell et al., (2013) and O'Connor et al. (2020), a sample of 10–25% of data units would be sufficient. Therefore, the second author randomly selected 20% (n=8) articles to double-check the analysis results. The interrater reliability of Cohen's Kappa was as high as 0.92. Corrections were made after a thorough discussion of disagreements in the initial code. Thematic analysis was then conducted, during which codes and categories were derived from the 42 studies to identify emerging themes relevant to the research questions. Multiple rounds of coding and recoding were carried out through the collaborative efforts of the authors. These phases of independent and joint analyses were designed to ensure the trustworthiness of the findings.

To triangulate the data, the study used the Structural Topic Model (STM) in R for topic modeling (Roberts et al., 2014). STM uses machine learning algorithms to identify key features of a corpus (Grajzl & Murrell, 2019). For the topic modeling, the abstract of each reviewed article was processed, including converting all text to lowercase, removing stop words and punctuation, and applying word stemming. The model generated ten topics using K-metrics and the Gibbs method, which are described in detail below. The Wordcloud2 package was used to visualize the most frequent keywords in the reviewed articles.

## Results

### *What is the current state of research publications on the use of LLMs in computer programming education?*

#### *Journal or proceeding publication outlets*

Of the reviewed articles, 21 were published in 2023 and 21 in 2024. Moreover, 17 are journal articles, and 25 are from conference proceedings. These publications appeared in 19 different journals or conference proceedings. Most articles were from proceedings of *ACM TSiCSE* (n=14) and *SIGCSE* (n=7). Additionally, journal articles appeared in various publications such as *Education Sciences* (n=2), *Computers & Education: Open* (n=1), *Computers in Human Behavior: Artificial Humans* (n=1), etc.

*Collaborations in LLMs for computer programming education research*

Thirty-eight out of the 42 studies involved author collaboration (see Figure 2). Most collaborators came from the same institution (n=22), while others involved multiple institutions: two institutions (n=10), three institutions (n=7), four institutions (n=2), and five institutions (n=1). Additionally, most collaborations took place within the same country (n=30), while international collaborations involved partnerships across two (n=5), three (n=5), or four (n=2) different countries.

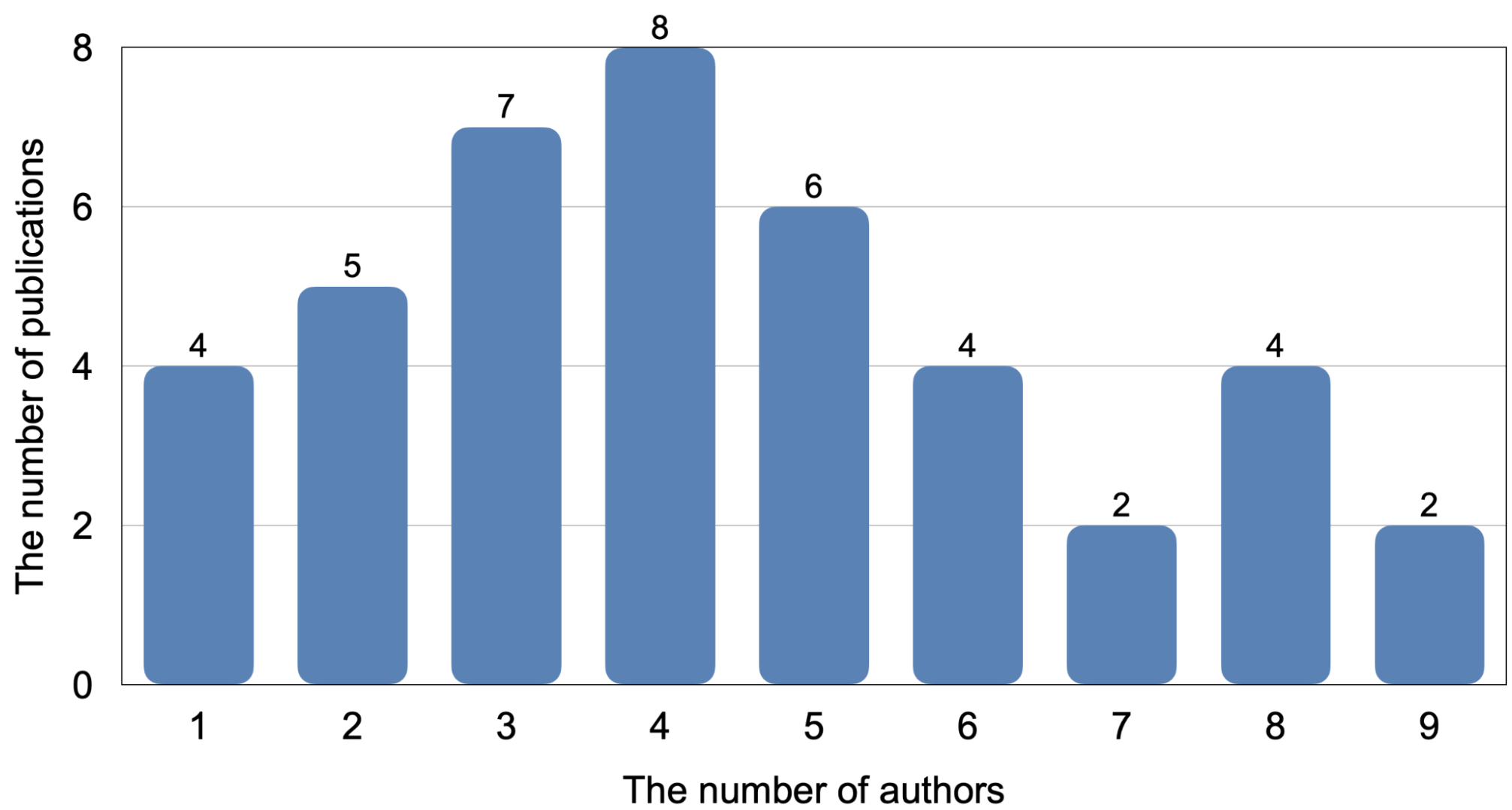

**Fig. 2.** The number of author collaborations in the reviewed publications

Regarding the authors' academic backgrounds, 39 out of the 42 articles included authors from Computer Science, Engineering, Computer and Information Sciences, or Human-Computer Interaction disciplines. Only two studies involved authors exclusively from the field of education (i.e., Hartley et al., 2024; Sun et al., 2024), and two articles included collaborative work between education scholars and scholars from STEM-related fields (i.e., Jing et al., 2024; Prather et al., 2023).

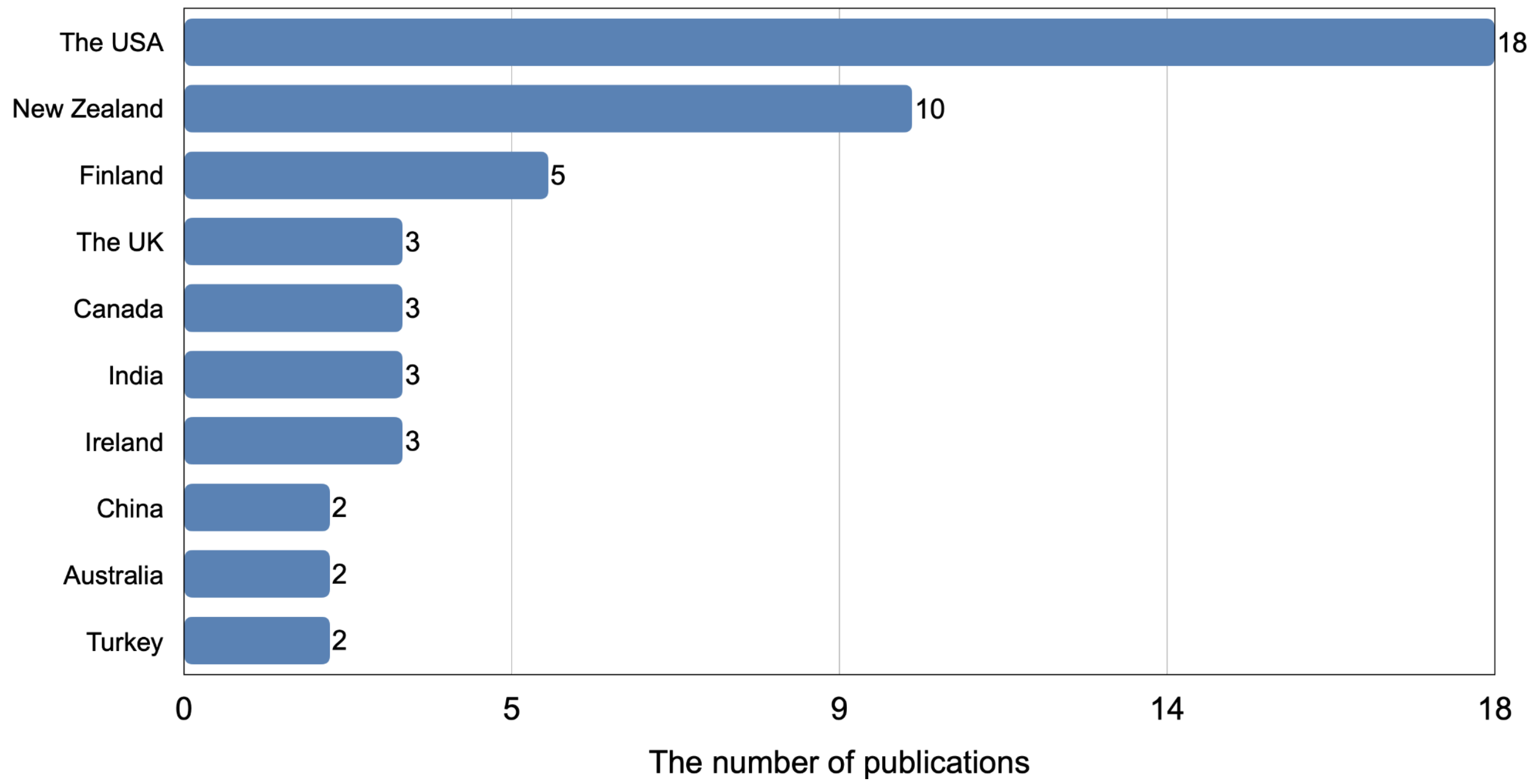

**Fig. 3.** Countries of the authors' affiliations (Note: if one article has several authors from the same country, this country will be counted once)

A total of 185 authors contributed to the 42 reviewed studies. They are from four continents: North America (e.g., USA and Canada), Oceania (e.g., New Zealand and Australia), Europe (e.g., Finland, UK, Ireland, Czech Republic, Portugal, Romania, Slovenia, Sweden), and Asia (e.g., China, India, Turkey, Indonesia, Israel, Japan, Jordan, Singapore, United Arab Emirates), as well as South America (e.g., Brazil). Since Turkey is geographically located at the crossroads of Asia and Europe, it was counted in both regions. The top three countries contributing authors to LLMs research in computer programming education were the USA (n=18), New Zealand (n=10), and Finland (n=5), followed by the UK (n=3), Canada (n=3), India (n=3) and Ireland (n=3) (see Figure 3). While in terms of where the study was conducted, the USA exceeded other countries with 12 studies, followed by New Zealand (n=4), Finland (n=3), and the UK (n=3).

*Research design and methods*

Out of the 42 reviewed articles, 16 focused on exploring and evaluating the performance of LLM tools. Case study design was used in 14 articles, followed by experimental design in three articles, survey studies in three, interview studies in two, and quasi-experimental design in two (see Figure 4). The research methods used in these studies were quantitative (n=22), qualitative (n=11), and mixed methods (n=9).

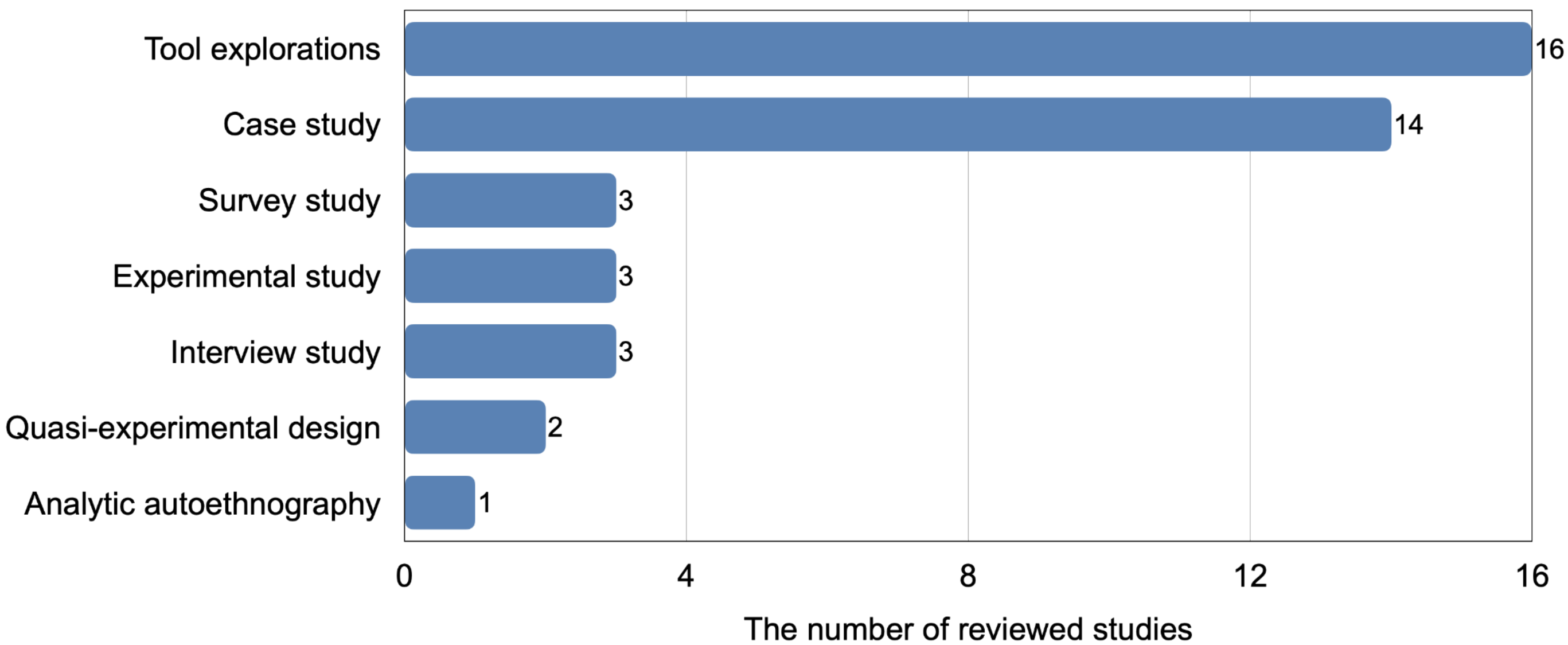

**Fig. 4.** Research design

The data sources in these studies included documents such as user interactions with LLM tools (n=22), surveys (n=14), assignments (n=7), interviews (n=5), exams (n=4), exercises (n=2), log data (n=2), and pre-and post-tests (n=2) (see Figure 5). For data analysis, researchers employed descriptive statistics (n=30), thematic analysis (n=12), inferential statistics (n=9), content analysis (n=2), and machine learning (n=1). The sample sizes in these studies varied from six to 1,211 participants, with an average of 25 participants per study.

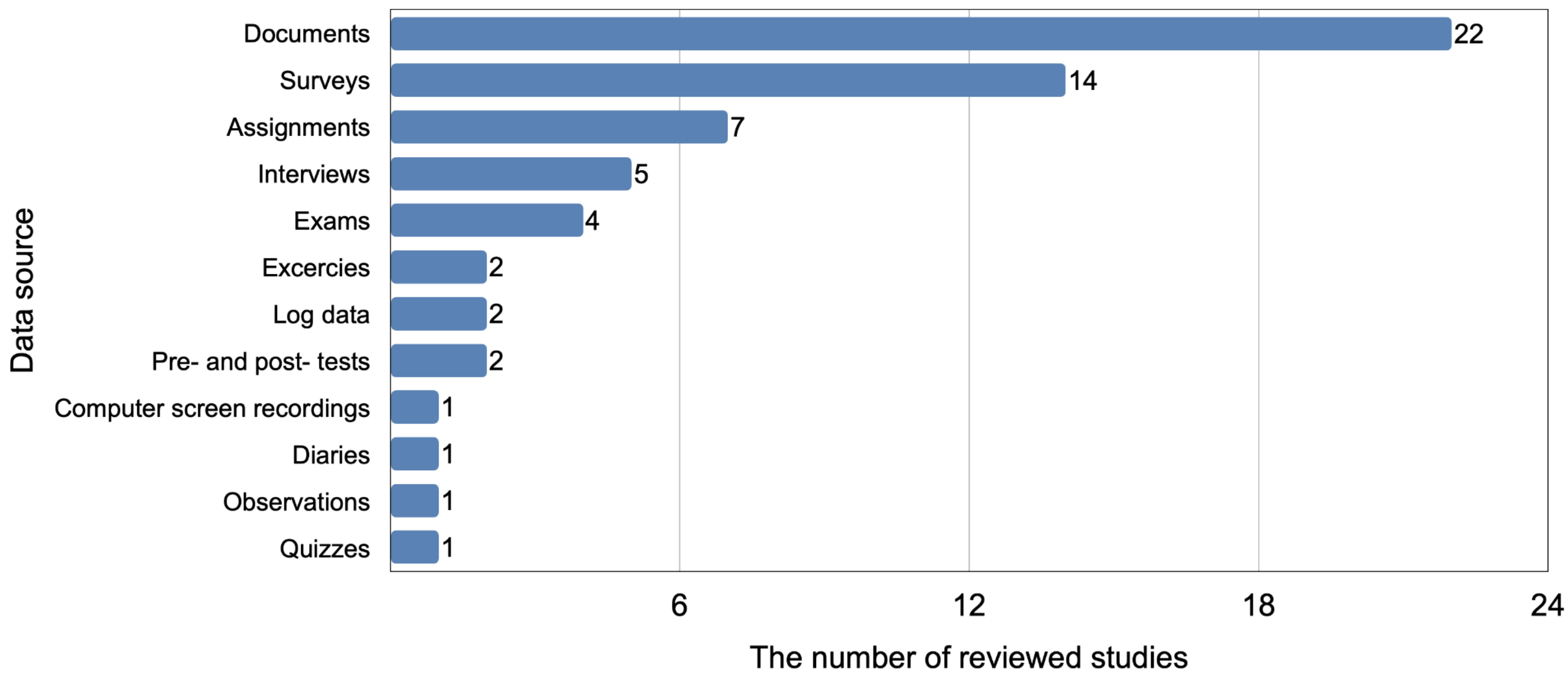

**Fig. 5.** Data sources (Note: some articles may have more than one data source)

The studies reviewed were primarily conducted in courses such as general programming and computer science (n=10), introductory programming (n=8), Python programming (n=7), object-oriented programming (n=6), data structures and algorithms (DSA) (n=1), web software development (n=1), etc. Twenty of the 42 articles focused on Python programming language, while others involved Java (n=4), C (n=4), C++ (n=2), C# (n=1), JavaScript (n=2), and Scala (n=1). The teaching strategies emphasized in these studies were predominantly hands-on learning and project-based learning (n=10), along with scaffolding (n=3) and self-directed or self-regulated learning (n=4). The LLM tools used were primarily ChatGPT (n=34), followed by GitHub Copilot (n=2), Bard (n=1), DCC-help (n=1), CS50 AI (n=1), Promptly (n=1), Refactor (n=1), and auto-explanation embedded in an ebook (n=1).

Regarding delivery mode, 19 studies were conducted in a traditional face-to-face setting, three were conducted online, and one was conducted in flipped classrooms. The educational setting was not reported in the remaining studies. Most of the research focused on undergraduate-level education (n=34), with a smaller portion involving graduate-level education (n=3).

**Table 2** A summary of the time duration of interventions

| Time duration of education interventions | n | Articles |
|---|---|---|
| One semester | 6 | French et al. (2023), Ishizue et al. (2024), Karnalim et al. (2023), Kosar et al. (2024), Liu et al. (2024), Sheard et al. (2024) |
| 3-12 weeks | 8 | Humble et al. (2023), MacNeil et al. (2023), Shoufan (2023), Silva et al. (2024), Sun et al. (2024), Taylor et al. (2024), Yilmaz & Yilmaz (2023, a, b) |
| 30 minutes - 80 minute | 2 | Prather et al. (2023), Jing et al. (2024) |

Of the 42 studies, only 16 reported educational interventions with students. The duration of these interventions varied from 30 minutes (n=1) to one semester (n=6) (see Table 2). Eight studies lasted between 3 and 12 weeks, while another four studies spanned an entire semester. Additionally, two studies involved brief interventions lasting between 30 and 80 minutes.

*Research topics based on keywords and abstracts*

A word cloud of keywords from the 42 articles showed that "ChatGPT," "large language models," and "artificial intelligence" were the most prominent keywords (see Figure 6). These were followed by terms such as "code generation," "assessment," and "academic dishonesty" in the context of "introductory programming" courses.

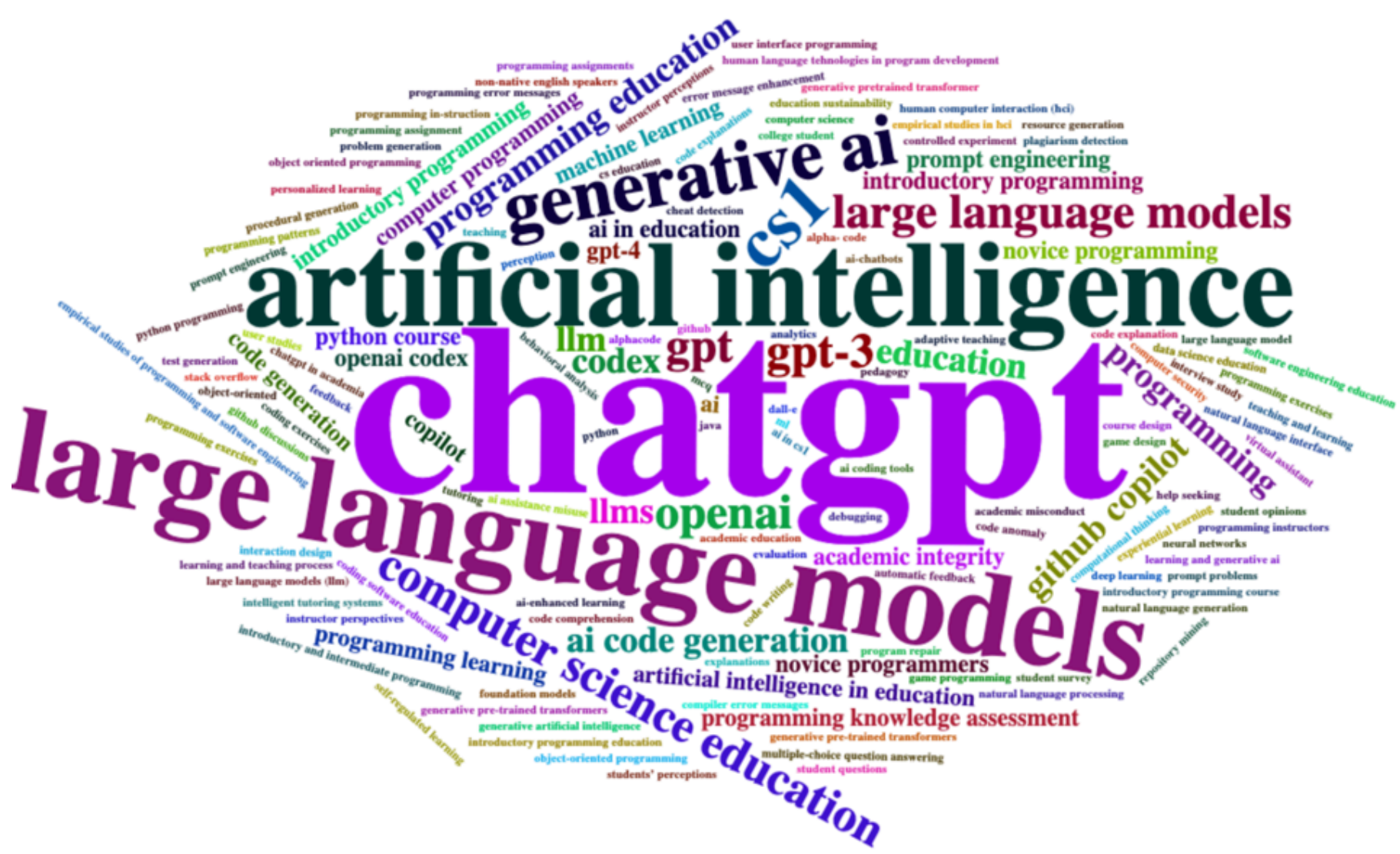

**Fig. 6.** Wordcloud of the keywords from the 42 reviewed studies

The research employed STM on the abstracts of the 42 articles, revealing the following ten topics (see Table 3). These findings concur with the thematic analysis results detailed in the sections below.

**Table 3** Top 10 topics based on STM from the abstract of the 42 reviewed studies

| Topic No | Topic Contribution of Tokens | Relevant terms | Main Themes |
|---|---|---|---|
| 1 | 17.9% | "chatgpt," "student," "learn," "tool," "use," "can," "educ," "code," "studi," "program" | Students use LLMs as learning tools for coding and programming education |
| 2 | 13.8% | "program," "code," "model," "assess," "llms," "cours," "gpt," "student," "provid," "use" | Assessing students' programming and coding skills using LLMs |
| 3 | 13.0% | "prompt," "problem," "chatgpt," "code," "assign," "tool," "student," "perform," "model," "solv" | Prompt engineering in utilizing ChatGPT to solve coding assignments |
| 4 | 12.0% | "exercis," "program," "problem," "code," "solut," "languag," "model," "generat," "student," "chatgpt" | Students leverage LLMs to generate solutions for programming problems in exercises |
| 5 | 8.9% | "program," "chatgpt," "use," "research," "student," "studi," "technolog," "learn," "percept," "instruct" | Student and instructor perceptions of using ChatGPT for programming education |
| 6 | 7.5% | "student," "program," "tool," "genai," "result," "generat," "educ," "assign," "trust," "like" | Students' trust in using generative AI tools to generate assignment results |
| 7 | 7.4% | "code," "program," "model," "use," "submiss," "student," "correct," "repair," "detector," "copilot" | Instructors detect programming assignment submissions using LLMs |
| 8 | 7.2% | "chatgpt," "program," "educ," "signific," "use," "technolog," "studi," "learn," "question," "potenti" | The significance and potential benefits of using ChatGPT for programming education |
| 9 | 7.1% | "code," "explan," "student," "larg," "generat," "help," "explain," "error," "llms," "doc" | Students use the code explanation function of LLMs to explain code errors |
| 10 | 5.1% | "tool," "code," "instructor," "plan," "program," "studi," "comput," "adapt," "research," "generat" | Instructors adapt LLMs as tools for creating computer programming lesson plans |

***How are the LLMs being used in computer programming education?***

*Instructor use*

LLMs can support instructors with various aspects of instructional design, teaching, and assessment. For instructional design and teaching, LLMs can assist instructors in creating lesson plans (e.g., Humble et al., 2023), generating content materials such as exercise banks and personalized tasks (e.g., Sheard et al., 2024), teaching programming lab sessions (e.g., Yilmaz & Yilmaz, 2023a, b), serving as a teaching assistant (e.g., Liu & Mhiri, 2024), and handling administrative tasks (e.g., Liu et al., 2024). For instance, Yilmaz and Yilmaz (2023a) incorporated ChatGPT into programming laboratory sessions in flipped classrooms. In their study, students in the experimental group used ChatGPT to solve object-oriented programming problems, while instructors used UML visuals to present the problems to enhance computational thinking and prevent direct reliance on ChatGPT for solutions. Liu and Mhiri (2024) found LLMs useful as teaching assistants to help students understand coding concepts and complex topics and provide guidance through hints rather than offering direct solutions for homework. Similarly, Harvard University researchers developed an AI tool for teaching CS50 Introduction to Computer Science that supported students' cognitive learning and answered curricular and administrative queries, thereby saving instructors' time (e.g., Liu et al., 2024).

Moreover, LLMs can also support instructors in assessing students' learning by generating assignments and examination questions (e.g., Humble et al., 2023), particularly in offering targeted feedback to learners in summative assessments (e.g., Liu et al., 2024; Sheard et al., 2024). In Sheard et al. (2024), the authors interviewed 12 instructors and a computer science instructor interviewee suggested using an automated system to provide feedback to students when they seek help.

*Learner use*

For learners' LLMs use, the reviewed articles revealed that LLMs can support learning in several aspects, such as programming project design (e.g., Sun et al., 2024), code generation (e.g., Prather et al., 2023; Sheard et al., 2024; Singh, 2023; Sun et al., 2024), code explanation (e.g., Silva et al., 2024; Singh, 2023), code debugging (e.g., Prather et al., 2023; Silva et al., 2024; Sun et al., 2024; Taylor et al., 2024), code optimization and refactoring (e.g., Ishizue et al., 2024; Kosar et al., 2024), understanding concepts (e.g., Roger et al., 2024), documentation and code examples (e.g., Sheard et al., 2024; Silva et al., 2023), getting feedback (e.g., Taylor et al., 2024), writing reports (e.g., Singh, 2023), solving project problems (e.g., Denny et al., 2024; French et al., 2023; Jing et al., 2024), and preparing for quizzes and assignments (e.g., Roger et al., 2024; Singh et al., 2023). LLMs can essentially serve as personal tutors for learners (e.g., Shoufan, 2023). For instance, Sun et al. (2024) conducted a quasi-experimental study in which students in the experimental group used ChatGPT to assist with class projects. They discovered that students utilized ChatGPT to design projects, decompose problems, acquire unfamiliar knowledge, write code, and debug. Prather et al. (2023) found that students benefited from using GitHub Copilot for their homework assignments, including code generation and suggestions, which helped their study process. Taylor et al. (2024) utilized "dcc-help," an LLM tool, to scaffold students' learning by providing feedback on code debugging. Similarly, Shoufan (2023) had students use ChatGPT as a tutor to ask questions about concepts, code completion tasks, and code analysis.

***What are the benefits of using LLMs in computer programming education?***

*Benefits for instructors*

LLMs can help instructors save time (e.g., Husain, 2024; Lau & Guo, 2023; Malinka et al., 2023; Sheard et al., 2024) by streamlining the preparation of learning and assessment tasks, providing feedback to students (e.g., Hellas et al., 2023; Liu & Mhiri, 2024; Sheard et al., 2024), and minimizing the time needed for office hours (e.g., Liu & Mhiri, 2024; Taylor et al., 2024). Additionally, LLMs can detect AI-generated code, as demonstrated by Karnalim et al. (2024), who developed a simple detector for AI-assisted code to aid instructors.

Husain (2024) conducted interviews with 12 programming instructors and found that ChatGPT improved programming instructors' efficiency and reduced their workload. It offered new instructional practices, customized to students' individual learning needs, assisted in performance assessment, and suggested lesson plans and teaching strategies. In a similar study, Lau and Guo (2023) interviewed 20 introductory programming instructors from 9 different countries and discovered that instructors recommended the long-term integration of AI tools in current courses to personalize students' learning and save instructors' time.

Regarding feedback, Liu and Mhiri (2024) used LLMs as virtual teaching assistants and found that these virtual TAs outperformed human TAs in terms of clarity and engagement. Similarly, Hellas et al. (2023) examined LLMs' responses to novice programmers' requests for help and discovered that LLMs provided effective feedback to learners, particularly in matters related to program logic, though they struggled with output formatting.

*Benefits for learners*

The reviewed studies demonstrate that using LLMs can positively benefit learners' cognitive learning (e.g., French et al., 2023; Humble et al., 2023; Jing et al., 2024; Ouh et al., 2023; Prather et al., 2023; Shoufan, 2023; Yilmaz & Yilmaz, 2023a, b), affective learning (e.g., Prather et al., 2023; Singh et al., 2023; Shoufan, 2023; Sun et al., 2024; Yilmaz & Yilmaz, 2023a, b), as well as the overall learning experience (e.g., Leinonen et al., 2023; Liu et al., 2024; MacNeil et al., 2023; Malinka et al., 2023; Silva et al., 2024; Taylor et al., 2024).

For instance, Yilmaz and Yilmaz (2023a) discovered that LLMs can enhance students' computational thinking skills, programming self-efficacy, and learning motivations. In a separate study, Yilmaz and Yilmaz (2023b) surveyed students on using ChatGPT for programming and found that students believe ChatGPT supports cognitive learning and self-confidence and helps the learning process by improving thinking skills and facilitating debugging. Additionally, Prather et al. (2023) found that utilizing GitHub Copilot can boost students' cognitive coding, metacognition, positive emotions, and learning efficiency.

Moreover, the use of ChatGPT can significantly influence students' learning experiences and processes. For instance, Silva et al. (2024) examined Brazilian students' perceptions of using ChatGPT and found that its integration into coding and programming courses enhances students' perceptions of educational support and personalized learning experiences. Liu et al. (2024) discovered that when students used an LLM virtual assistant in programming courses, they found the AI tool to be helpful, effective, and reliable.

*No significant differences in learning outcomes*

However, some research has shown that there are no significant differences in programming learning outcomes between students who use ChatGPT and those who do not (e.g., Kosar et al., 2024; Sun et al., 2024). For instance, Kosar et al. (2024) conducted an experimental study and found no significant difference in learning outcomes between groups, although students perceived that learning with ChatGPT was better than learning without it. Similarly, Sun et al. (2024) conducted a quasi-experimental study in a programming course to investigate the effects of ChatGPT-supported programming on students' performance, behaviors, and perceptions. They found no significant difference between the experimental group (using ChatGPT) and the control group (not using ChatGPT). However, they did observe a slight improvement in learning performance among students who used ChatGPT for programming.

***What are the limitations and concerns regarding using LLMs in computer programming education?***

*Limitations of LLMs*

The primary limitations of LLMs identified in the reviewed studies include inaccuracy (e.g., Amoozadeh et al., 2024; Balse et al., 2023; Denny et al., 2024; Hellas et al., 2023; Humble et al., 2023; Husain, 2024; Kosar et al., 2024; Joshi et al., 2024; Liu et al., 2024; Ouh et al., 2023; Prather et al., 2023; Roger et al., 2024; Sheard et al., 2024; Shen et al., 2024; Shoufan, 2023; Sun et al., 2024; Yilmaz & Yilmaz, 2023 b; Wermelinger, 2024), undesirable responses (e.g., Husain, 2024; Taylor et al., 2024), limited input and output capabilities (e.g., Sun et al., 2024), language barriers (e.g., Humble et al., 2023), text-only constraints (e.g., Ouh et al., 2023), limited visual generation capabilities (e.g., French et al., 2023), and bias (e.g., Joshi et al., 2024; Prather et al., 2023).

Inaccuracy is the most mentioned limitation of LLMs. Humble et al. (2023) discovered that the accuracy of LLM responses varies depending on the prompt, and the accuracy in languages other than English is lower due to the training data being primarily in English. Instructors interviewed by Husain (2024) expressed concerns about ChatGPT potentially generating inaccurate information, especially when prompts are too general or lack context. Balse et al. (2023) highlighted that students cannot directly use LLMs for feedback due to accuracy issues.

Another main limitation is LLMs' undesirable responses. Taylor et al. (2024) noted that LLMs might disregard prompts and provide undesirable responses. Hellas et al. (2023) observed LLMs' unreliability, particularly when formatting output as required by automated assessment systems. Ouh et al. (2023) mentioned that some LLMs are limited to text, which is a significant limitation. However, advancements in ChatGPT4 and DALL-E are gradually reducing this constraint. Moreover, the bias in LLMs is one of the important limitations. It often originates from training datasets containing harmful stereotypes related to gender, race, emotion, class, and other characteristics (Prather et al., 2023).

*Concerns of using LLMs in programming education*

The concerns regarding the use of LLMs in programming education include academic dishonesty (e.g., French et al., 2023; Hellas et al., 2023; Humble et al., 2023; Husain, 2024; Karnalim et al., 2024; Lau & Guo, 2023; Malinka et al., 2023; Prather et al., 2023; Roger et al., 2024; Sheard et al., 2024; Singh, 2023; Yilmaz & Yilmaz, 2023 a, b), impeded learning, including potential setbacks to creative thinking, problem-solving skills, and communication skills (e.g., Denny et al., 2024; French et al., 2023; Husain, 2024; Kosar et al., 2024; Leinonen et al., 2023; Prather et al., 2023; Savelka et al., 2023; Silva et al., 2024; Singh, 2023; Yilmaz & Yilmaz, 2023b), data privacy (e.g., Husain, 2024), copyright issues (e.g., Humble et al., 2023; Lau & Guo, 2023; Prather et al., 2023), occupational concerns (e.g., Prather et al., 2023; Shoufan, 2023), diversity and equity challenges (e.g., French et al., 2023; Hellas et al., 2023; Lau

& Guo, 2023), time consumption (e.g., Prather et al., 2023), unstructured learning (e.g., Husain, 2024), and negative affect (e.g., Denny et al., 2024).

Academic dishonesty is a major concern. Researchers pointed out that instructors and students worry about the potential misuse of LLMs for cheating. In an interview with programming instructors, Husain (2024) found concerns about students relying on easily accessible solutions from LLMs. Humble et al. (2023) also highlighted the risk of cheating with LLMs. To decrease cheating risks, researchers have proposed various strategies, such as using paper-based assessments (e.g., Lau & Guo, 2023), incorporating visual representations of problems (e.g., Yilmaz & Yilmaz, 2023 a), emphasizing formative assessments and the learning process, and adopting varied assessment formats such as oral presentations and explanations, and context-specific problems (e.g., Lau & Guo, 2023). For instance, Yilmaz & Yilmaz (2023 a) addressed academic honesty concerns by adjusting assignments with visual UML diagrams to deter students from copying assignment descriptions into LLMs.

Another concern has been raised regarding whether LLMs genuinely enhance learning or potentially hinder cognitive and affective learning. Kosar et al. (2024) mentioned concerns about the potential negative impact of LLMs on students' cognitive and communication skills. Studies by Yilmaz & Yilmaz (2023b) and Savelka et al. (2023) raised concerns that over-reliance on LLMs might hinder their critical thinking abilities. Similarly, Husain (2024) found that programming instructors worried that ChatGPT might adversely affect students' critical thinking, problem-solving, programming, and human interaction skills, all of which are essential for student development. Students reported that they worried about LLMs affecting their creativity and ability to write code (Denny et al., 2024) or using GitHub Copilot could distract them during class since they relied on Copilot as a safety net (e.g., Prather et al., 2023).

Regarding data privacy and security, researchers were worried about the risk of exposing students' private information to LLMs. For instance, Husain (2024) found that programming instructors were concerned about the possibility of students' personal data being exposed to ChatGPT. In terms of copyright and intellectual property, Prather et al. (2023) recommended that educators teach students how to properly utilize LLMs and understand how these models are trained to prevent illegal or unethical use of the technology.

Several studies suggest that students may have concerns about the impact of using LLMs on their future careers (e.g., Prather et al., 2023; Shoufan, 2023). In Prather et al. (2023), student interviewees expressed that GitHub Copilot's ability to write code could potentially replace software developers. Shoufan (2023) found that a significant portion of participants feared ChatGPT might take over human coding jobs, while others viewed ChatGPT as a complementary resource.

Furthermore, LLMs might contribute to societal inequalities. French et al. (2023) highlighted concerns about concentrating knowledge and power in the hands of a few elite individuals who control powerful supercomputers, potentially leading to job loss due to AI. Lau and Guo (2023) revealed instructors' concerns about the potential for LLMs to accelerate inequity by disadvantaging those who are unfamiliar with AI tools. Jordan et al. (2024) noted the language and culture differences regarding GPT's functions.

Time consumption depends on how students use the LLMs. For novice learners who are not familiar with using LLMs and programming, using LLMs may lead them to spend more time on programming and debugging. Prather et al. (2023) observed that students using LLMs for coding sometimes exhibited "drifting" behaviors, where they spent significant time deciphering or debugging LLM suggestions, causing them to get off track and waste time.

Moreover, using LLMs may also cause negative effects and emotions. For instance, Denny et al. (2024) mentioned that using LLMs could occasionally harm students' self-esteem, as they may spend considerable time trying to solve a problem while LLMs can provide a solution in seconds.

### ***What implications does current research have on future research and practice on LLMs for computer programming education?***

*Implications for future research*

Regarding domains, researchers encourage future studies to explore the impact of using LLMs not only on cognitive learning but also on students' communication skills (e.g., Kosar et al., 2024), engagement, collaboration, critical thinking, and creativity (e.g., Husain, 2024). They also recommend examining the reasons behind students' behaviors when using LLMs for programming learning (e.g., Prather et al., 2023). Additionally, future research should not only focus on learners but also explore LLMs' influence on educators (e.g., Kosar et al., 2024).

Regarding LLMs' integration into education, more research is needed to identify the most effective ways to integrate LLMs into computer programming education (e.g., Savelka et al., 2023; Silva et al., 2024; Singh, 2023) and assess the extent of the impact of using ChatGPT on students' performance (e.g., Singh, 2023). Specific research might also focus on prompt engineering to help users generate appropriate prompts for LLMs (e.g., Denny et al., 2024; Jing et al., 2024; Shoufan, 2023; Silva et al., 2024; Sun et al., 2024; Taylor et al., 2024). Roger et al. (2024) suggested exploring how the use of ChatGPT affects students' educational trajectories. More specifically, future research could explore when and how to provide feedback to novice programming learners (e.g., Hellas et al., 2023), scaffolding novice learners' understanding, tailoring AI coding tools for pedagogy, adapting IDEs for AI-aware teaching, equity and accessibility, efficacy studies, advanced computing courses, scaling instruction, and rethinking foundational courses in light of AI tools (e.g., Lau & Guo, 2023).

As for assessment research, although AI can be useful for auto-grading to save instructors' time, it should be approached cautiously, taking ethical considerations into account (e.g., Lau & Guo, 2023). For academic dishonesty, Hoq et al. (2024) proposed to explore what types of AI use constitute academic dishonesty. Roger et al. (2024) suggested future studies explore whether learners' trust in AI influences their likelihood of using AI for cheating. Moreover, to detect academic dishonesty using LLMs, researchers should investigate AI features and tools that can help detect AI-assisted submissions (e.g., Karnalim et al., 2024; Lau & Guo, 2023; Savelka et al., 2023). Additionally, As AI tools evolve, research into appropriate assessment methods remains essential (e.g., Savelka et al., 2023). Besides pedagogical explorations, Lau and Guo (2023) recommended that future research concentrate on developing theories on effectively utilizing AI tools for learning.

With respect to research methods, researchers are encouraged to employ varied data sources, such as interaction data and learning outcomes, as well as methods like mixed methods research (e.g., Humble et al., 2023; Jing et al., 2024). For research design, experimental studies have been proposed to better understand the factors that improve the integration of ChatGPT in programming education (e.g., Husain, 2024). Additionally, most current research spans only a limited time frame; therefore, more longitudinal studies are needed to explore these phenomena (e.g., Prather et al., 2023; Sun et al., 2024; Yilmaz & Yilmaz, 2023 a, b). Researchers also aim for larger sample sizes (e.g., Humble et al., 2023; Ishizue et al., 2024) across various programming courses and languages, such as Java, covering diverse educational levels (e.g.,

Balse et al., 2023). Future studies should also include different programming tasks with varying difficulty levels (e.g., Cipriano, 2023; Denny et al., 2024; Gutierrez et al., 2024; Jing et al., 2014; Karnalim et al., 2024; Kosar et al., 2024; Reeves et al., 2023; Shoufan, 2023; Yilmaz & Yilmaz, 2023 a, b).

*Implications for future education practice*

The studies reviewed offered several suggestions for educational practices, such as integrating LLMs in computer programming education, enhancing AI literacy through prompt engineering, addressing the ethics of LLM use in education, designing improved LLM tools for learning (e.g., Prather et al., 2023), and providing students with guidelines and advice (e.g., Silva et al., 2024; Singh, 2023). Additionally, researchers emphasized the importance of learning programming skills despite using LLMs (e.g., Piccolo et al., 2023; Silva et al., 2024).

To effectively integrate LLMs into education, researchers suggested that education practitioners reevaluate learning goals (e.g., Sheard et al., 2024) and develop new curricula, in-class activities, teaching strategies, and course outlines to efficiently incorporate LLMs into programming curricula (e.g., Gutierrez et al., 2024; Husain, 2024; Sheard et al., 2024; Sun et al., 2024). Educators can design questions and problems that promote students' critical thinking, higher-order cognitive skills, and motivation (e.g., Gutierrez et al., 2024; Joshi et al., 2024; Savelka et al., 2023). Additionally, using LLMs for monitoring student learning and providing automatic formative feedback and grading can be beneficial (e.g., Hoq et al., 2024; Ishizue et al., 2024; Liu et al., 2024). Liu et al. (2024) proposed the development of AI auto-grading tools to reduce instructors' grading workload. To decrease the negative effects of using LLMs, researchers recommend emphasizing critical thinking and revising assessment strategies to prevent academic dishonesty (e.g., Husain, 2024; Kosar et al., 2024; Lau & Guo, 2023; Ouh et al., 2023; Piccolo et al., 2023; Popovici et al., 2024; Sheard et al., 2024; Shoufan, 2023; Silva et al., 2024; Sun et al., 2024). Educators can implement authentic assessments (e.g., Shoufan, 2023), including paper exams and oral, video, or image-based evaluations (e.g., Lau & Guo, 2023). They can also create student-specific questions about students' code and provide personalized feedback (e.g., Piccolo et al., 2023; Silva et al., 2024). Another strategy is to require students to explain the code they submit (Popovici et al., 2024) and increase in-class assessments (Roger et al., 2024).

Researchers emphasized the importance of enhancing AI literacy for both instructors and students, including the critical skill of prompt engineering (e.g., Jing et al., 2024; Roger et al., 2024; Sun et al., 2024; Yilmaz & Yilmaz, 2023 a, b) and AI ethics (e.g., Hoq et al., 2024; Prather et al., 2023). Educators should be aware of the benefits and drawbacks of using LLMs in programming education (e.g., Sun et al., 2024; Wermelinger, 2024). With sufficient AI literacy, educators can carefully integrate AI-based resources into programming education (e.g., Sun et al., 2024) and guide students to understand AI's advantages and limitations (e.g., Jing et al., 2024; Wermelinger, 2024). Teaching students prompt engineering is particularly important in AI literacy (Shen et al., 2024). Roger et al. (2024) proposed that instructors provide a set of prompts that students can input into ChatGPT to generate a wide range of practice questions. For instance, educators were encouraged to incorporate metacognitive prompts to support students' self-regulated learning (e.g., Yilmaz & Yilmaz, 2023 a). Additionally, instructors can develop tools to assist learners in crafting effective prompts for LLMs (e.g., Denny et al., 2024). Educating students about ethical considerations is also essential to AI literacy (e.g., Hoq et al., 2024; Prather et al., 2023). Computer programming educators should understand and define the appropriate level of trust students should place in Generative AI tools (e.g., Amoozadeh et al., 2024).

Researchers advocate for the customization of LLM AI tools for programming education (e.g., Hellas et al., 2023; Prather et al., 2023). For example, Prather et al. (2023) recommended improving GitHub Copilot tools to enhance student learning by allowing learners more control over prompts, offering shorter suggestions, providing scaffolding, using explainable AI, and addressing data copyright concerns.

Moreover, one effective approach to integrating LLMs into programming education is to establish clear guidelines for instructors and students on the use of LLMs in computer programming education (e.g., Roger et al., 2024; Silva et al., 2024; Singh, 2023). Roger et al. (2024) recommended that educators set explicit rules for each course or assignment, providing detailed instructions on ChatGPT use and specifying acceptable practices.

## Discussion and implications

This study presents a systematic review of LLMs in computer programming education in higher education settings. The goal is to enhance the understanding of the current status of research in the field, including methods of use of LLMs, benefits, limitations, concerns, and future research and practice implications. The paper selection process followed a four-phase PRISMA framework (Page et al., 2021). We reviewed 42 empirical studies published in peer-reviewed journals and three CS education focused conference proceedings from 2023 to March 2024. Data were analyzed using bibliometric analyses, thematic analysis, and topic modeling. First, we analyzed basic bibliometric data to summarize the current state of publications, research contexts, participant demographics, sample sizes, etc. Next, we synthesized detailed strategies for using LLMs, as well as the benefits, limitations, concerns, and implications for future research and practice from the 42 reviewed studies.

***LLM in computer programming education research***

This review study found that 16 out of 42 studies concentrate on the initial stage of exploring and evaluating the performance of LLM tools. However, research on the impact of LLMs on various learning domains is limited. Future studies could examine a wider range of learning domains, such as cognitive learning, affective learning, and overall learning experiences (Husain, 2024; Kosar et al., 2024). Additionally, research is needed on when and how to integrate LLMs into computer programming education, as well as effective strategies for assessing learning outcomes.

Of the reviewed studies, only 16 have educational interventions in academic settings, with five employing experimental or quasi-experimental methods. All study interventions lasted one semester or less, reflecting the limited research approaches, as noted in previous research on AI for computer science education (Authors, in review). A broader range of research approaches (Authors, 2023; Zhang & Aslan, 2022), including more experimental or quasi-experimental studies with longitudinal data (Husain, 2024), is necessary to better understand the impact of LLMs on programming education and integrate LLMs into education practices.

The findings of this study indicated that despite the fact that most studies involve collaborations, most authors of the reviewed studies come from CS and informatics disciplines. Only two studies involve researchers from education, and another two involve collaboration between CS and education experts. Research indicated that expanding researchers' networks is needed to enhance our understanding of a phenomenon and research quality (Gorska et al., 2020). Future research in this field could benefit from incorporating diverse expertise from multiple disciplines, such as CS and education, to deepen our understanding of the phenomenon and enhance educational practices.

### *Education practice*

The findings highlighted the importance of integrating LLMs into programming education and the significance of AI and LLM literacy. Based on the results of this study, we propose a conceptual framework to guide educators in integrating LLMs into computer programming education. The framework includes stakeholders (who), objectives (why), and specific activities (how) and highlights the importance of LLM literacy and use guidelines (see Figure 7).

First, we recommend that educators identify the primary stakeholders, such as instructors or students, involved in integrating LLMs into computer programming education.

Second, practitioners should establish clear objectives for integrating LLMs, whether to assist instructors with course and instruction design, teaching, and assessment or to support students' learning outcomes (e.g., cognitive and affective) and learning processes.

The third step is related to determining how to achieve these objectives guided by educational pedagogy, whether behaviorism, cognitivism, or constructivism. For instruction design, instructors can leverage LLMs to design courses, lesson plans, and resources. For teaching, LLMs can provide examples and act as teaching assistants during in-class activities such as laboratory sessions. For assessment, LLMs can generate exercises, exam questions, and problems and detect cheating.

Regarding supporting learners' cognitive learning, students can use LLMs to generate project ideas, code, explanations, debugging, refactoring, and documentation. LLMs can also enhance students' affective learning by increasing engagement, motivation, self-efficacy, and interest in programming. Additionally, LLMs can improve students' overall learning experience by offering ease of use, increased learning efficiency (Husain, 2024; Prather et al., 2023), and enjoyable learning journeys (Liu et al., 2024; Malinka et al., 2023; Silva et al., 2024; Taylor et al., 2024).

Moreover, LLM literacy is important for computer programming instructors and students throughout the process. This literacy includes an understanding of LLM concepts, applications, prompt engineering, limitations, and ethical considerations. Regarding LLM concepts, instructors and students should have a foundational understanding of LLMs and associated techniques. In terms of applications, instructors should know how to use LLMs in design, teaching, and assessment, and learners should know how to leverage LLMs to support learning effectively. Prompt engineering involves crafting precise and effective prompts for LLMs to obtain desired outputs. Additionally, instructors and students should recognize the limitations of LLMs and understand their boundaries. Instructors and students must also make informed decisions regarding the ethical use of LLMs, including considerations such as academic integrity, data privacy, and copyright. These AI literacy skills are essential for equipping instructors and students with the knowledge and capabilities needed to efficiently and effectively integrate LLMs in computer programming education. To ensure instructors and students use LLMs appropriately and effectively, it is crucial to establish and share clear guidelines on when and how to use LLMs in computer programming education. These guidelines should offer practical advice on incorporating LLMs into teaching and learning processes.

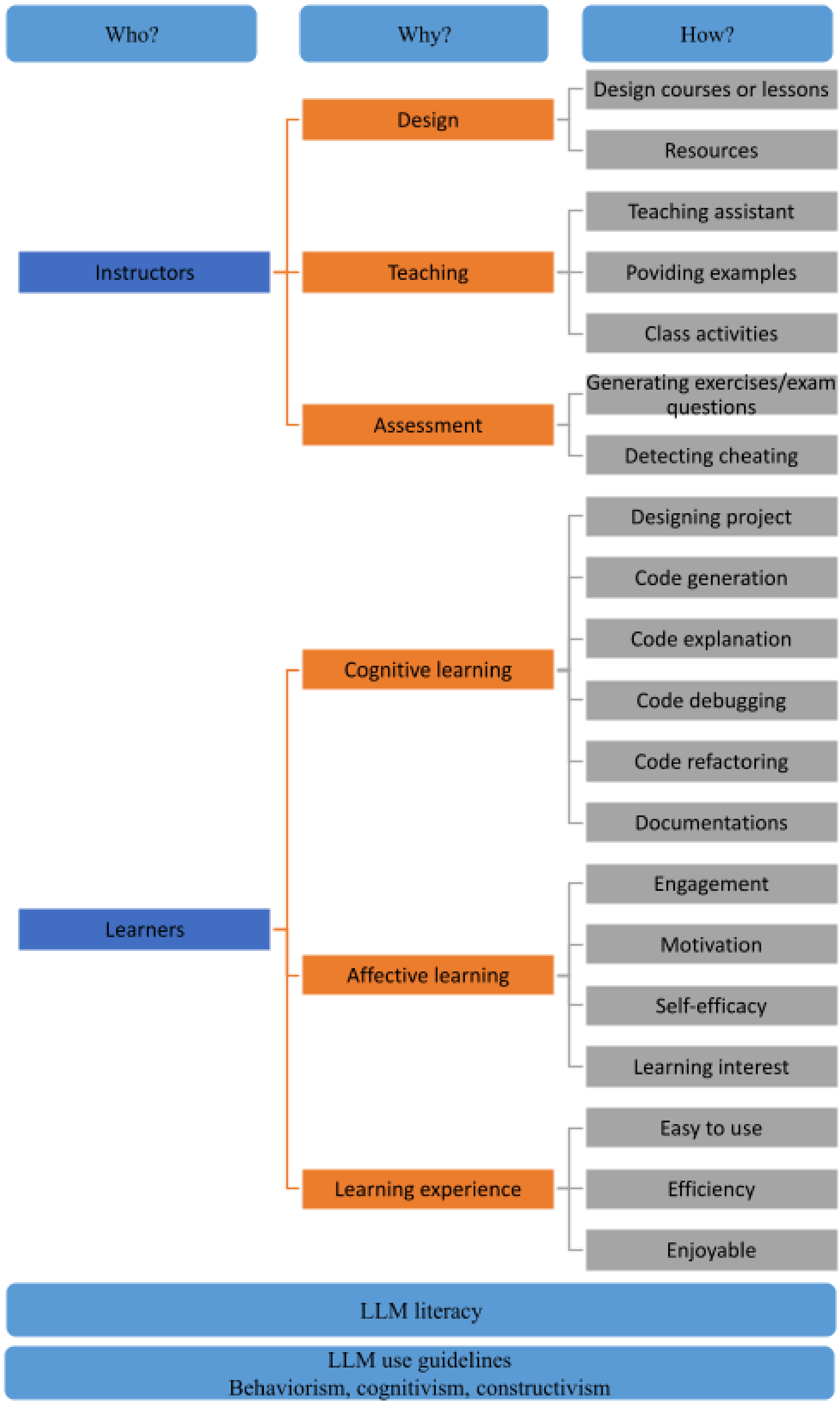

Who?
Why?
How?
Instructors
Design
Design courses or lessons
Resources
Teaching
Teaching assistant
Poviding examples
Class activities
Assessment
Generating exercises/exam questions
Detecting cheating
Learners
Cognitive learning
Designing project
Code generation
Code explanation
Code debugging
Code refactoring
Documentations
Affective learning
Engagement
Motivation
Self-efficacy
Learning interest
Learning experience
Easy to use
Efficiency
Enjoyable
LLM literacy
LLM use guidelines
Behaviorism, cognitivism, constructivism

**Fig. 7.** The conceptual framework to guide educators in integrating LLMs into computer programming education.

## Limitations and future research

This systematic review has some limitations that need to be recognized. First, this review study focused on peer-reviewed journal articles and three computer programming education-oriented conference proceedings, adhering to specific inclusion and exclusion criteria throughout several rounds of searching and selecting. Only three specific conference proceedings (i.e., SIGCSE, ICER, and ITiCSE) were included; other international or regional conference proceedings, such as Australasian Computing Education Conference, Koli Calling International Conference, and CHI, were not included. The intentional selection of peer-reviewed journal articles and conference proceedings can ensure the quality of the reviewed studies to some extent. It is possible that valuable studies published as book chapters, reports, or dissertations were excluded from this review. Future research could benefit from incorporating a wider variety of publication types.

Second, only English-language publications were considered in this review study, which led to the exclusion of research in other languages, such as Chinese or Spanish. Future reviews could expand the scope to include research published in different languages for a more global perspective on using LLMs in computer programming education.

Third, due to their focus on education, the review study primarily used databases like Web of Science (SCI/SSCI), SCOPUS, and EBSCOhost. Research publications indexed in other databases, such as IEEE Xplore or ACM digital library, were not included. Future reviews could enhance their comprehensiveness by expanding the range of databases used in literature searches.

Since the release of GPT-3.5 in November 2022, there has been increased attention from researchers and practitioners on the use of LLMs in computer programming education. Therefore, this review specifically focused on articles published between 2023 and March 2024 to capture the latest trends and research in the field. While there may have been earlier research on using LLMs in computer programming before 2023, and newer research could emerge after March 2024 as the technology continues to evolve, future reviews could extend the time frame to analyze the broader trends in using LLMs in computer programming education.

## Conclusions

In this systematic review, we used bibliometric analyses, thematic analysis, and topic modeling to analyze and synthesize data. This comprehensive approach facilitated data triangulation and provided a systematic overview of how LLMs are currently being integrated into computer programming education, along with their benefits, limitations, concerns, and future implications for research and practice. We also proposed a conceptual framework to guide instructional designers and instructors in effectively integrating LLMs within higher education computer programming education.

This systematic review offers valuable insights into the role of LLMs in computer programming teaching and learning, contributing to our understanding of how these tools can enhance educational processes in higher education. LLMs can be successfully integrated into computer programming education by following the steps and considerations outlined in the framework. The study’s findings can also inform future empirical research on LLMs and computer programming education and guide future systematic reviews on this topic.

**References** * Refers to the studies reviewed in this systematic review study